\documentclass[prl,amsmath,amssymb,aps,superscriptaddress,floatfix,reprint,notitlepage]{revtex4-1}
\pdfoutput=1
\usepackage{natbib}
\usepackage{graphicx}
\usepackage{color}
\usepackage{dcolumn}
\usepackage{bm}
\usepackage[utf8]{inputenc}
\usepackage{amssymb}
\usepackage{color,amsmath, braket}
\usepackage{mathtools, ulem}
\usepackage{soul,xcolor} %use \st{text} to achieve strikethrough
\setstcolor{red} %set colour of strikethrough to red
\usepackage{epstopdf}

\DeclareUnicodeCharacter{03BD}{{$\nu$}}

\usepackage{hyperref}
\hypersetup{
 pdfnewwindow=true, colorlinks=true,
 linkcolor=blue, anchorcolor=blue,
 citecolor=blue, filecolor=blue,
 menucolor=blue, urlcolor=blue}

\begin{document}

\title{Experimental determination of the energy per particle in partially filled Landau levels}% Force line breaks with \\
% \thanks{A footnote to the article title}%

\author{Fangyuan Yang}
\affiliation{Department of Physics, University of California at Santa Barbara, Santa Barbara CA 93106, USA}
\author{Alexander A. Zibrov}
\affiliation{Department of Physics, University of California at Santa Barbara, Santa Barbara CA 93106, USA}
\author{Ruiheng Bai}
\affiliation{Department of Physics, University of California at Santa Barbara, Santa Barbara CA 93106, USA}
\author{Takashi Taniguchi}
\affiliation{International Center for Materials Nanoarchitectonics,
National Institute for Materials Science,  1-1 Namiki, Tsukuba 305-0044, Japan}
\author{Kenji Watanabe}
\affiliation{Research Center for Functional Materials,
National Institute for Materials Science, 1-1 Namiki, Tsukuba 305-0044, Japan}
% \author{Mark O. Goerbig}
% \affiliation{Laboratoire de Physique des Solides, CNRS UMR 8502, Universit\'e Paris-Saclay, 91405 Orsay Cedex, France}
\author{Michael P. Zaletel}
\affiliation{Department of Physics, University of California, Berkeley, CA 94720 USA}
\author{Andrea F. Young}
\email{andrea@physics.ucsb.edu}
\affiliation{Department of Physics, University of California at Santa Barbara, Santa Barbara CA 93106, USA}

%\noaffiliation

\date{\today}

\begin{abstract}
We describe an experimental technique to measure the chemical potential, $\mu$, in atomically thin layered materials with high sensitivity and in the static limit.  We apply the technique to a high quality graphene monolayer to map out the evolution of $\mu$ with carrier density throughout the N=0 and N=1 Landau levels at high magnetic field.  
%We observe sequences of jumps in $\mu$ at fractional Landau level fillings, $\nu$, associated with  2- and 4-flux composite fermion fractional quantum Hall gaps, and broad regions of negative $\partial \mu/\partial \nu$ near integer $\nu$ associated with Wigner crystallization of charge carriers. 
By integrating $\mu$ over filling factor, $\nu$, we obtain the ground state energy per particle, which can be directly compared with numerical calculations.
In the N=0 Landau level, our data show exceptional agreement with numerical calculations over the whole Landau level without adjustable parameters, as long as the screening of the Coulomb interaction by the filled Landau levels is accounted for. In the N=1 Landau level, comparison between experimental and numerical data reveals the importance of valley anisotropic interactions and the presence of valley-textured electron solids near odd filling. 
\end{abstract}

\maketitle

Partially filled Landau levels (LLs) are a paradigmatic example of flat band systems where dominant Coulomb interactions lead to a rich phase diagram of correlation driven electron states. 
Theoretically, the partially filled LL provides a compromise between phenomenological richness and computational tractability. However, quantitatively benchmarking numerical methods with transport measurements is typically limited to a discrete set of LL filling factors, $\nu$.  
Thermodynamic quantities such as the chemical potential $\mu$ are more closely related to theoretically calculable quantities. 
Owing to recent progress in improving sample quality\cite{dean_fractional_2020} and the fact that the single particle band structure is known to a high degree of accuracy, graphene is an ideal venue to pursue quantitative understanding of partially filled LLs.
In this Letter we report precise measurements of $\mu$ in a high quality monolayer graphene layer at both zero and high magnetic fields. 
Typical measurements of thermodynamic quantities in graphene probe the compressibility $\partial n/\partial \mu$  at finite  frequency\cite{martin_observation_2008,feldman_unconventional_2012,feldman_fractional_2013,zibrov_even-denominator_2018}, hindering accurate measurements in the quantum Hall regime where equilibration times can become long. Our measurements probe $\mu$ directly\cite{lee_chemical_2014} in the static, $\omega\rightarrow0$ limit. This allows us to determine $\mu$ across a continuous range of $\nu$, and subsequently the total energy per flux quantum, $E$, where $\mu=\partial E/\partial \nu$.

\begin{figure*}[t!]
\includegraphics{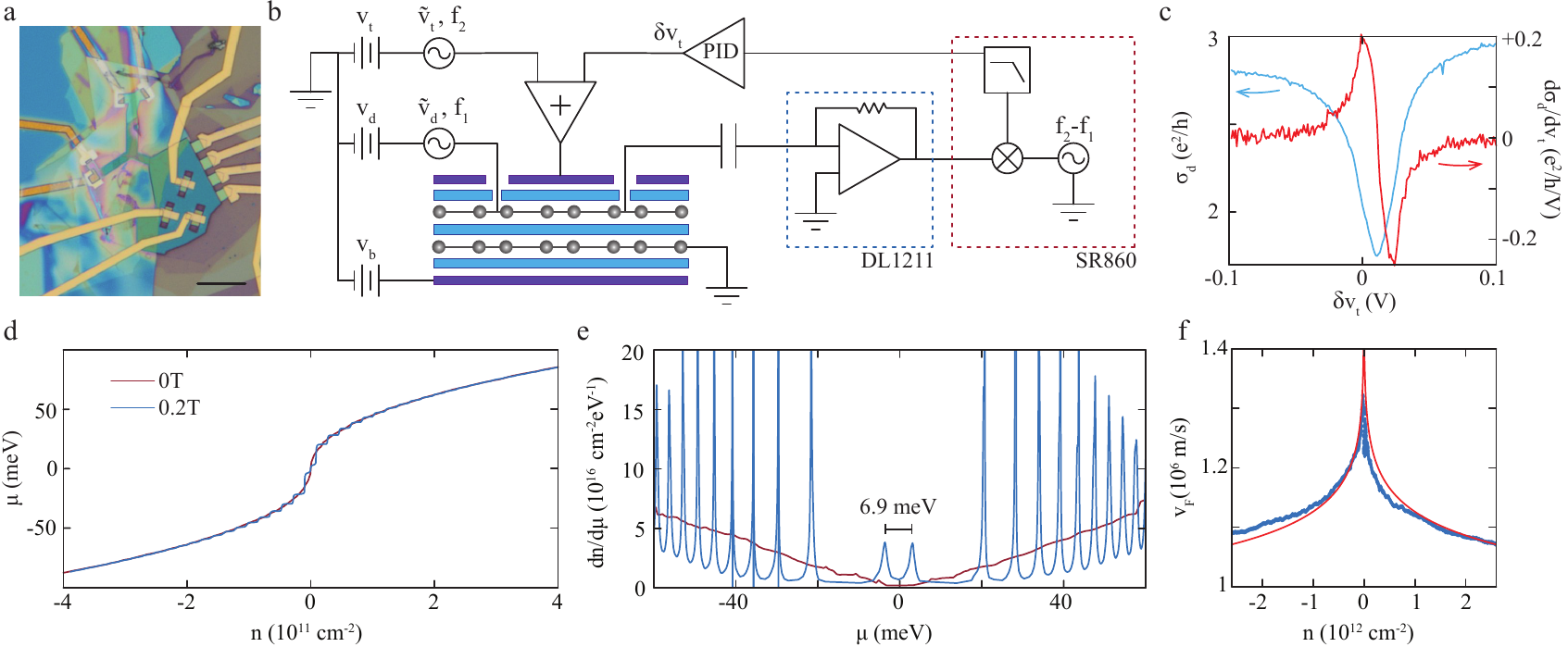}
\caption{\label{fig:figure1} 
\textbf{(a)} Optical image of the device. The scale bar is 10$\mu$m.
\textbf{(b)} Measurement schematic.  Static gate voltages are applied to the top gate ($v_t$), bottom gate ($v_b$), and detector monolayer ($v_d$).  AC voltages are applied to one detector layer contact at  $f_1 = 13.77$Hz ($\tilde v_d$) and to the top gate at $f_2 = 110$Hz ($\tilde v_t$), producing a current proportional to $\sigma_d$ at $f_1$ and to $d{\sigma_d}/dv_{t}$ at $f_2-f_1$ at the second detector layer contact, measured with a DL1211 current preamplifier and demodulated with a SR860 lock-in amplifier.  
$d{\sigma_d}/dv_{t}$ serves as the error signal for a digital feedback loop (PID) whose output ${\delta}v_{t}$ is added to the top gate voltage, fixing the carrier density of the detector. Under these conditions, ${\delta}{\mu}=-c_t{\delta}v_{t}/c_0 $.
\textbf{(c)} ${\sigma_d}$ (blue) and ${\delta}{\sigma_d}/{\delta}v_{t}$ (red) as a function of ${\delta}v_{t}$. 
\textbf{(d)} $\mu(n)$ at B=0T (red) and $0.2$T (blue), measured at T=15mK. \textbf{(e)} Density of states $dn/d{\mu}$ calculated by numerical differentiation of data in panel (d). 
The ZLL is split by a sublattice gap\cite{hunt_massive_2013,amet_insulating_2013} of $\Delta_{AB}=6.9$meV. 
\textbf{(f)} $n$-dependent $v_F$ measured by fitting B=0T data to ${\mu}^2=({\Delta_{AB}}/2)^2+({\hbar}v_F\sqrt{{\pi}|n|})^2$ with $\Delta_{AB}$ fixed and $v_F$ a free function of $n$. The red curve is a fit to theoretical models\cite{gonzalez_marginal-fermi-liquid_1999,das_sarma_many-body_2007,polini_graphene_2007} of Fermi velocity renormalization by Coulomb interactions.
}
\end{figure*}

Our heterostructure consists of two  graphene monolayers embedded between top and bottom graphite gates (see Figs. \ref{fig:figure1}a-b and \ref{fig:SI_device}), with each conducting layer separated by a hexagonal boron nitride (hBN) dielectric of approximately 40nm thickness.  
The dual graphite-gated structure ensures low charge inhomogeneity on both graphene monolayers while allowing independent control of their respective carrier densities through the static gate voltages applied to the top gate ($v_t$), bottom gate ($v_b$), and top monolayer ($v_d$). Internal contacts\cite{yan_charge_2010,zhao_magnetoresistance_2012,zhu_edge_2017,polshyn_quantitative_2018,zeng_high-quality_2019} are attached to the top monolayer---designated the `detector'---and are used to measure its bulk conductivity $\sigma_d$. The charge density of the detector layer is $n_d=c_t(v_t-v_d)+c_0(\phi-v_d)$. Here $c_t$ and $c_0$ are the top gate-detector and detector-sample geometric capacitances and $\phi$ is the electric potential of the sample monolayer.
% \mz{This discussion has a sign error}
To measure $\mu$, we ground the sample layer so that $\phi=\mu$, and keep $v_d$ constant.  
Variations in $\mu$ are then given by ${\delta}{\mu}= \frac{1}{c_0} \delta n_d -\frac{c_t}{c_0} {\delta}v_{t}.$ Next, we adjust $\delta v_t$ to maintain $\delta n_d = 0$\cite{eisenstein_negative_1992}.  This gives $\delta\mu$ via the simple relation ${\delta}{\mu} =-{c_t}{\delta}v_{t}/{c_0}$, with the only input being the capacitive lever arm $c_t/c_0$, which can be precisely measured (see Fig. \ref{fig:SI_CapLeverArm}).

Functionally, $\delta n_d = 0$ is enforced by choosing a ``target'' density $n_d$ such that $\sigma_d$ is at a conductance minimum corresponding to the Dirac point at B=0T or a weak FQH state at high B.  Figure 1b shows the schematic of our measurement circuit.  $\sigma_d$ is measured in voltage bias mode, by applying an AC voltage $\tilde v_d$ at frequency $f_1$ to one of the internal contacts and measuring the resulting current.  $\sigma_d$ measured at B=0T is shown in Fig. \ref{fig:figure1}c. 
In order to mitigate the effects of contact resistance in the detector, which are also tuned by $v_t$ and $v_d$, we use $\frac{d \sigma_d}{d v_t} = 0$ as the feedback condition. 
To do so, we apply an additional voltage modulation to the top gate ($\tilde v_t$) at frequency $f_2$. Demodulating the current at frequency $f_2-f_1$ produces a signal proportional to $\frac{d \sigma_d}{d v_t}$. The value of $\delta v_t$ is then adjusted by a feedback loop to zero this signal, giving the desired $\delta \mu$.
While the current measurement is done at finite frequency to allow low noise readout, it does not require charging of the sample layer at these frequencies.  This allows us to access regimes where the sample layer conductivity is very small and equilibration times are very large.  In practice, measurements are typically done with equilibration times of $\tau\approx 1$~sec.

Fig.~\ref{fig:figure1}d shows $\mu$ measured at B=0T and 200mT, plotted as a function of the sample carrier density $n=c_0(v_d-\mu)+c_b(v_b-\mu)$, where $c_b$ is the capacitance between the sample and the bottom gate.
$\mu(n)$  shows the $\sqrt{n}$ dependence expected for the linearly dispersing bands of monolayer graphene\cite{martin_observation_2008}, as well as steps associated with LL formation when a small magnetic field is applied.  
To quantitatively model the data, we take ${\mu}^2=({\Delta_{AB}}/2)^2+({\hbar}v_F\sqrt{{\pi}|n|})^2$, where $\Delta_{AB}$ is the sublattice splitting\cite{hunt_massive_2013,amet_insulating_2013} and $v_F$ is the Fermi velocity. 
We determine $\Delta_{AB}=6.9$meV from the splitting of the zero energy LL (ZLL) centered at $\mu=0$, evident in Fig. \ref{fig:figure1}e where we plot $dn(\mu)/d\mu$ as determined by numerical differentiation of the $\mu(n)$ data (see also Fig. \ref{fig:SI_LFDOS}).
Figure \ref{fig:figure1}f shows $v_F(n)$, determined by fixing $\Delta_{AB}$ but allowing $v_F$ to be a free $n$-dependent parameter.  
$v_F$ is enhanced at low densities, consistent with past experiments\cite{elias_dirac_2011,chae_renormalization_2012} and well fit by theoretical models of Fermi velocity renormalization\cite{gonzalez_marginal-fermi-liquid_1999,das_sarma_many-body_2007,polini_graphene_2007}, as shown by  the red curve in Fig. \ref{fig:figure1}f and described in the SI. 

\begin{figure}[t]
\includegraphics[width=\columnwidth]{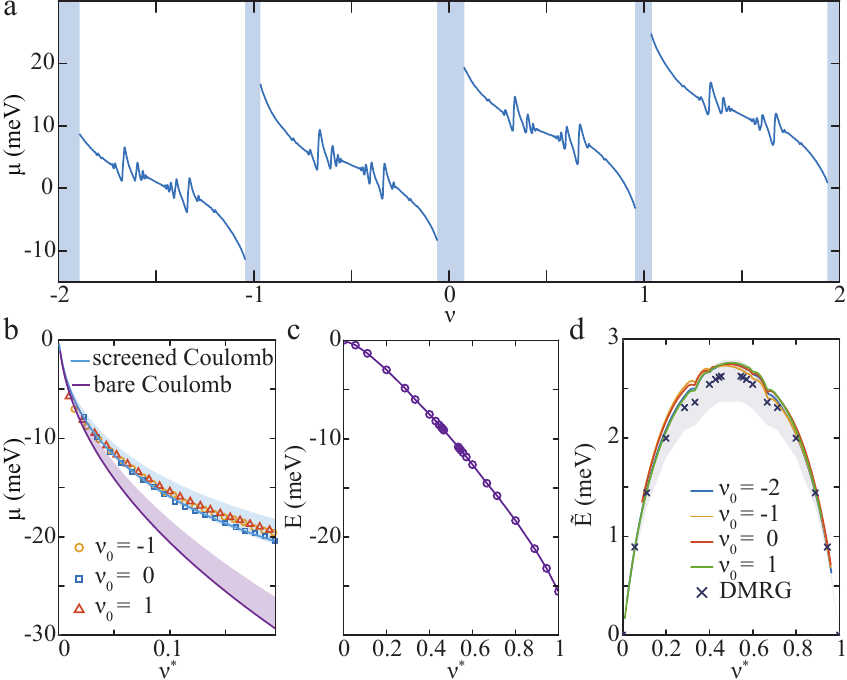}
\caption{\label{fig:figure2} %Chemical potential in the zero energy LL.
\textbf{(a)} $\mu(\nu)$ within the ZLL measured at B=14T and nominal T=15mK. Blue regions indicate domains  of $\nu$ where the charging time of the sample exceeds the measurement time of $\sim$1 second. (see Fig. \ref{fig:SI_hysteresis}). 
\textbf{(b)}  $\mu$ at B=18T and nominal T=40mK for low $\nu^*$, measured relative to $\nu = -1$ (orange), $\nu = 0$ (blue), and $\nu = 1$ (red).  
The cyan and purple curves are calculated $\mu$ for a Wigner crystal with screened and unscreened Coulomb interactions, respectively, taking $\epsilon_{\mathrm{hBN}}=4.0$ and $\alpha_G=1.85$; shaded ranges reflect uncertainty in those parameters as described in main text.  
\textbf{(c)} Numerically calculated\cite{zaletel_infinite_2015} total ground state energy of the N=0 LL after accounting for the screened Coulomb interactions. 
\textbf{(d)} Comparison of experimentally determined (solid lines) and numerically calculated (dark blue crosses) $\tilde E$. 
Both experimental and numerical data have a linear-in-$\nu^*$ background subtracted so that $\tilde E$ vanishes at integer $\nu^*$. Data were taken at B=18T and T=40mK.}
\end{figure}

At high magnetic fields, the LLs of monolayer graphene are approximately four-fold degenerate due to the spin and valley degrees of freedom.  
Fig.~\ref{fig:figure2}a presents $\mu(\nu)$ at B=14T across the ZLL that spans $-2<\nu<+2$, where $\nu=2\pi\ell_B^2n$ is the LL filling factor.
The high quality of the detector layer is crucial for achieving high experimental $\mu$ resolution, as FQH conductivity minima in the detector layer provide sensitive transducers for the sample layer chemical potential (see Fig. \ref{fig:SI_Corbino}). 
Over large regions of density, $\mu(\nu)$ decreases as a function of $\nu$ (negative compressibility), despite the naive expectation that $\mu$ should increase monotonically with $\nu$ due to Coulomb repulsion. This is because the chemical potential measured here is actually relative to that of a classical capacitor, which subtracts off the $q=0$ part of the Coulomb interaction $\frac{1}{2} V(q=0) n^2$. It is well understood \cite{fano_configuration-interaction_1986, eisenstein_negative_1992} that negative compressibility then arises because correlations lower the energy of quantum Hall states relative to that of a uniform charge distribution. $\mu$ jumps at each integer $\nu$ indicating incompressible integer quantum Hall states arising from the broken symmetry of the spin and valley components of the isospin.
Additional jumps are observed at a series of fractional $\nu$ associated with incompressible fractional quantum Hall (FQH) states at $\nu^*=p/2p\pm1$ ($p=1,2,3,...$) and $\nu^*=p/4p\pm1$ (with $p=1$ and $2$)\cite{eisenstein_compressibility_1994, feldman_unconventional_2012,feldman_fractional_2013,zibrov_even-denominator_2018}. 
Here $\nu^*=\left|\nu-\nu_0\right|$ indicates the filling relative to an adjacent integer filling $\nu_0\in \mathbf{Z}$.  
At high $B$, regions (shaded in blue) around integer $\nu$ are good insulators, and so are no longer accessible at low temperatures due to the hours- or days-long equilibration time of the sample layer (see Fig. \ref{fig:SI_hysteresis}). 

The four copies of the ZLL are nearly identical, suggesting that the LL is close to fully spin and valley polarized at this magnetic field.
This is expected based on the measured value of $\Delta_{AB}$, which splits the valley degree of freedom in the ZLL; in combination with the Zeeman energy, FQH physics is expected to be predominantly single component\cite{polshyn_quantitative_2018} in this regime of magnetic fields. We begin our quantitative analysis at low $\nu^*$ where electron Wigner crystal phases\cite{lam_liquid-solid_1984,levesque_crystallization_1984} are the expected ground state. 
In transport measurements, the Wigner crystal manifests as a low-temperature insulator that undergoes a metal-insulator transition at finite temperature due to pinning of the crystal by weak disorder, as observed in both GaAs/AlGaAs quantum wells\cite{goldman_evidence_1990} and more recently in graphene\cite{zhou_solids_2019}.
The largely classical nature of the correlations in this regime make thermodynamic modelling tractable, and quantitative agreement obtains between theory\cite{bonsall_static_1977} and compressibility measurements in GaAs/AlGaAs quantum wells\cite{eisenstein_negative_1992,eisenstein_compressibility_1994}. 

Fig.~\ref{fig:figure2}b shows $\mu$ plotted as a function of $\nu^*$ near different integer fillings within the ZLL.  For comparison, we also show theoretical calculations of $\mu$ in the Wigner crystal phase developed for the case of unscreened Coulomb interactions\cite{levesque_crystallization_1984}, where $\mu(\nu^*)=-1.173|\nu^*|^{1/2} E_C$.   Here $E_C=\frac{e^2}{\epsilon_{\mathrm{hBN}}\ell_B}$ is the Coulomb energy.
The model has only one parameter, the dielectric constant $\epsilon_{\mathrm{hBN}} = \sqrt{\epsilon^\parallel \epsilon^\perp}$, which is the geometric average of the in and out-of plane dielectric constants of the hBN substrate.
$\epsilon^\perp=3.0$ can be determined in situ, but $\epsilon^\parallel$ is not precisely known, though it is thought to be $\epsilon^\parallel \approx 6.6$\cite{geick_normal_1966}. 
Even accounting for uncertainty in this parameter,  the model does not agree with experiment.
Quantitative agreement is achieved, however, by considering the screening of the Coulomb interactions by the graphite gates, which are accounted for using standard electrostatic calculations, and by the filled Dirac sea, which we account for within the random phase approximation (RPA)\cite{shizuya_electromagnetic_2007}.  RPA takes as an additional input parameter the graphene fine structure constant $\alpha_G$. Still treating the electrons as a classical Wigner crystal, we numerically evaluate the Madelung-type energy for the screened interaction $V_{\textrm{scr}}(r)$ to obtain $\mu(\nu^\ast)$\cite{supp}.
To reflect uncertainty in the input parameters, we show a range spanning $\epsilon_{\mathrm{hBN}}\in(4.0,4.5)$ and $\alpha_G\in (1.75,2.2)$, in addition to reference curves for $\epsilon_{\mathrm{hBN}}=4.0$ and $\alpha_G=1.85$. 

The screened Coulomb interaction provides an exceptionally good match to the experimental data, suggesting that no additional effects are present and that accounting for the screening is sufficient to achieve quantitative understanding of this regime.  
We note that based on spin-wave transmission measurements\cite{zhou_solids_2019}, spin Skyrmions appear to play a role in the Wigner solid phases near $\nu=\pm1$. 
We do observe a small but systematic discrepancy between $\mu$ near even and odd integer $\nu$ in the Wigner crystal regime. This suggests that the large Zeeman energy, $E_Z\approx .03 E_C$, restricts the Skyrmion size to the point where they do not generate significant corrections to $\mu$ at low $\nu^*$.

\begin{figure}
\includegraphics[width=\columnwidth]{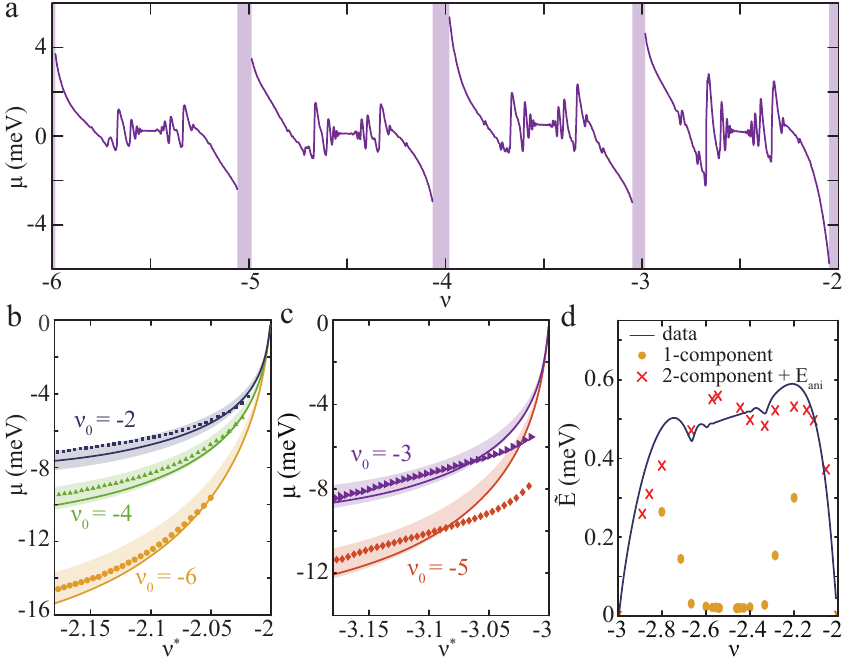}
\caption{\label{fig:figure3} 
\textbf{(a)} $\mu$ in the N=1 LL at T=15mK and B=13T. 
\textbf{(b)} $\mu$ measured near $\nu_0 = -2$, $-4$, and $-6$. Solid lines are $\mu$ calculated from the Wigner crystal model with parameters identical to those used in Fig.~\ref{fig:figure2}b. \textbf{(c)} $\mu$ near $\nu_0 = -3$ and $-5$. The solid lines showing the Wigner crystal model do not match the data, suggesting the importance of valley Merons\cite{cote_skyrme_2008} near these fillings. 
\textbf{(d)} 
Comparison of experimentally determined $\tilde E$ with numerical simulations  for $-3<\nu<-2$. 1-component numerical calculations underestimate  the experimental result by a significant margin. Including both valley components as well as the contribution of lattice scale anisotropies as in Eq. \ref{Eq:eani} with $g_z=g_{xy}=0.1 (a/\ell_B)E_C$ can restore agreement to within $100\mu eV\approx 2.5\times 10^{-3}E_C$.}
\end{figure}

Closer to the center of the LL, correlations become quantum in nature and even numerical calculation of $\mu$ is not tractable for arbitrary $\nu$. However, numerical methods can accurately calculate the \textit{total energy} per flux quantum $E(\nu)$ at many rational values of $\nu$, as has long been the focus of exact diagonalization and density matrix renormalization group (DMRG) studies.
Fig.~\ref{fig:figure2}c shows the ground state energy calculated using infinite DMRG\cite{zaletel_infinite_2015} (iDMRG) on a circumference $L = 18 \ell_B$ cylinder for a number of rational $\nu$, assuming wave functions are restricted to a single spin and valley component and making use of the screened interaction $V_{\textrm{scr}}$.

The  calculated $E$ is dominated by a linear background, $\mu_0 \nu^\ast$, that is proportional to the exchange-correlation energy of the integer quantum Hall effect; the correlations underlying the FQH effect are reflected in the deviations of the calculated $E$ from this background. 
In Fig. \ref{fig:figure2}d, we subtract off the linear contribution by instead plotting  $\tilde E =E -\nu^* E(\nu^*=1)$ (Fig.~\ref{fig:figure2}d), which ensures $\tilde E(0)=\tilde E(1)=0$. 
This can be compared with experiment by integrating $\mu(n)$, $\tilde E(\nu^\ast)= \int_0^{\nu^\ast}(\mu(\nu)-\mu_0) d\nu$, where $\mu_0$ is  chosen to ensure $\tilde E(0)=\tilde E(1)=0$.
To aid in fixing $\mu_0$ accurately, the experimental data is extrapolated to integer $\nu$ by using the Wigner crystal model. 
Numerical and experimental data  agree to within experimental uncertainty in $\alpha_G$ and $\epsilon_{\mathrm{hBN}}$ without additional adjustable parameters.  Similarly, the measured thermodynamic gap at charge neutrality, 53meV, agrees with theoretically calculated jump in $\mu$ to within 4\% \cite{supp}.  These constitute remarkably good quantitative agreement for a many-body system. 

Fig.~\ref{fig:figure3}a shows $\mu$ measured across the first excited LL, corresponding to orbital quantum number N=1 and spanning $\nu\in(-6,-2)$. In contrast to the N=0 level, both the size of the chemical potential jumps associated with FQH gaps\cite{polshyn_quantitative_2018} and the magnitude of the negative compressibility systematically decrease with increasing $|\nu|$. 
This trend arises naturally due to the nature of the screened Coulomb interaction $V_{\textrm{scr}}$ \cite{shizuya_electromagnetic_2007}: in the ZLL, particle-hole symmetry makes the screening $\nu$ independent, but within the N=1 LL screening smoothly interpolates between the N=0 and N=2 values as the four-component LL fills. Indeed, applying this interpolation to the Wigner crystal regime near even filling factors produces an excellent quantitative match between the data and theory (Fig.~\ref{fig:figure3}b).

The N=1 LL and ZLL are further distinguished by the effect of the sublattice symmetry breaking $\Delta_{AB}$, which splits the valleys in the ZLL but has negligible effect on the energies of the N=1 LL.  
This manifests most obviously in our data in the low-$\nu^*$ regimes around near \textit{odd} integer filling, shown in Fig.~\ref{fig:figure3}c. In contrast to the comparable regimes of $\nu^*$ near even integers, and throughout the ZLL, the data are not matched by the predictions of $V_{\textrm{scr}}$ for a single electron Wigner crystal.  To understand this data, we note that  tilted field magnetotransport experiments\cite{young_spin_2012} find evidence for a spin polarized state at $\nu=\pm4$ in which excitations are either single spin flips or small Skyrmions, similar to the situation at $\nu=\pm1$ in the ZLL. At $\nu=\pm3, \pm5$, in contrast, activated gaps show minimal tilted field dependence, consistent with the lowest energy charged excitations being valley textures.  Theoretically, the ground state of a spin-polarized but valley-unpolarized LL applicable to $\nu=\pm3, \pm5$ is then expected to be a solid of such valley textures\cite{cote_skyrme_2008}, with resulting corrections to $E$ and consequently to $\mu$.  Notably, the corrections to the energy will be largest when the valley textures are most extended.  The observed anomalous $\mu(\nu)$  supports the idea that the low single-particle valley anisotropy in the N=1 LL  stabilizes a solid of extended valley textures.  This could be tested in the future by extending  numerical calculations\cite{cote_skyrme_2008} of such solids to include the screened Coulomb interaction.    

The multicomponent nature of the N=1 LL is further evidenced in Fig.~\ref{fig:figure3}d, where iDMRG simulations of a \textit{single} component system fail to reproduce the experimentally determined $\tilde E$ when using the same model parameters which produce good agreement in the ZLL.
Interestingly, iDMRG finds a significantly lower total energy compared to experiment. This suggests a missing contribution to the energy, since adding degrees of freedom to a variational parameter space can only lower the numerically calculated energy, increasing the discrepancy.  
An appealing candidate is the anisotropy of the Coulomb interactions at small length scales, which breaks the valley-$SU(2)$ symmetry and can be expected to provide corrections of $E_{ani}\sim \frac{a}{\ell_B} E_C\approx 1.75$~meV at B=13T, where $a=.246$nm is the graphene lattice constant.  
Though known to be important in the ZLL\cite{dean_fractional_2020} near $\nu=0$, evidence for short range anisotropy in the N=1 LL has been limited to the observation of a possible valley-ordered state at $\nu=4$ for low magnetic fields\cite{polshyn_quantitative_2018}, and they have not received much attention in the theoretical literature\cite{alicea_graphene_2006,kharitonov_phase_2012}.

To model their effect, we analyze the interactions which arise when projecting a short-range Hubbard-$U$ interaction into the N=1 LL. For simplicity we assume full-spin polarization so that electrons are described by a two-component field $\psi_r$ indexed by valley $\tau^z$. It is convenient to express the result as the continuum interaction which would produce the same Hamiltonian if the electrons were in the N=0 LL.  
Taking into account the interplay of the form-factors of the N=1 LL and the sublattice structure, we find the general form\cite{supp}
\begin{align}
H_{\textrm{ani}} &=  \frac{1}{2} \int d^2 r_{1/2} \left[  g_{z} \psi^\dagger_{r_1} \tau^z \psi_{r_1} \ell_B^4 \nabla^{4} \delta(r_1 - r_2) \psi^\dagger_{r_2} \tau^z \psi_{r_2} \right . \notag  \\
& \left . \quad + g_{xy} \psi^\dagger_{r_1} \tau^x \psi_{r_1} \ell_B^2 \nabla^{2} \delta(r_1 - r_2) \psi^\dagger_{r_2} \tau^x \psi_{r_2} + (x \to y) \right ]\label{Eq:eani}
\end{align}
where $g_i \sim \frac{a}{\ell_B}E_c$.
Note that the interactions are \textit{derivatives} of $\delta$-functions; in contrast, the same exercise in the ZLL would find contact interactions\cite{kharitonov_phase_2012,sodemann_broken_2014}. 
Because the FQH effect around density $\nu^\ast = \frac{1}{m}$ attaches zeros
$(z_i - z_j)^m$ to the inter-electron wave function, a $\nabla^{2m} \delta$ interaction effectively ``turns-off'' for densities below $\frac{1}{m+1}$. In the ZLL, this means the anisotropies only operate for $-1 < \nu < 1$, while in the N=1 we predict the anisotropies act for all $2 + 1/3 < \nu < 6 - 1/3$.
This is indeed the region where our 1-component numerics deviate from experiment. 

Treating $g_z, g_{xy}$ as adjustable phenomenological parameters, we perform 2-component iDMRG numerics that include $H_{\textrm{ani}}$.  Fig.~\ref{fig:figure3}d shows the results for $g_{xy}=g_z = 0.1\frac{a}{\ell_B} E_C$, which agree with experiment to within 100 $\mu$eV, comparable to the discrepancies observed in the ZLL. In both LLs these discrepancies amount to $2\times10^{-3}$ of the bare Coulomb energy $E_C$.  

\begin{acknowledgments}
M.P.Z. acknowledges conversations with M. Ippoliti, Z. Papic, N. Regnault, and E. Rezayi, who generously provided exact-diagonalization energies, as well as M. Metlitski.
The iDMRG code used in this work was developed in collaboration with R. Mong and F. Pollmann. 
Experimental work by F.Y., A.A.Z., R.B and A.F.Y. was supported by the National Science Foundation under DMR-1654186. 
Work by M.P.Z. is supported by the Army Research Office under W911NF-17-1-0323. 
A portion of this work was performed at the National High Magnetic Field Laboratory, which is supported by the National Science Foundation Cooperative Agreement No. DMR-1644779 and the state of Florida. 
K.W. and T.T. acknowledge support from the Elemental Strategy Initiative conducted by the MEXT, Japan, Grant Number JPMXP0112101001,  JSPS KAKENHI Grant Number JP20H00354 and the CREST(JPMJCR15F3), JST.
A.F.Y. acknowledges the support of the David and Lucile Packard Foundation.
\end{acknowledgments}

\bibliographystyle{custom}
%\bibliography{references,supp}
%
 
\clearpage
%%%%%%%%%%%%%%%%%%%%%%%%%%%%%%%%%%%%%%%%%%%%%%%%%%%%
\pagebreak
\widetext
\begin{center}
\textbf{\large Supplementary Information}
\end{center}
\renewcommand{\thefigure}{S\arabic{figure}}
\renewcommand{\thesubsection}{S\arabic{subsection}}
\setcounter{secnumdepth}{2}
\renewcommand{\theequation}{S\arabic{equation}}
\renewcommand{\thetable}{S\arabic{table}}
\setcounter{figure}{0}
\setcounter{equation}{0}
\onecolumngrid

\section{Device fabrication method}

\begin{figure*}[ht!]
\includegraphics{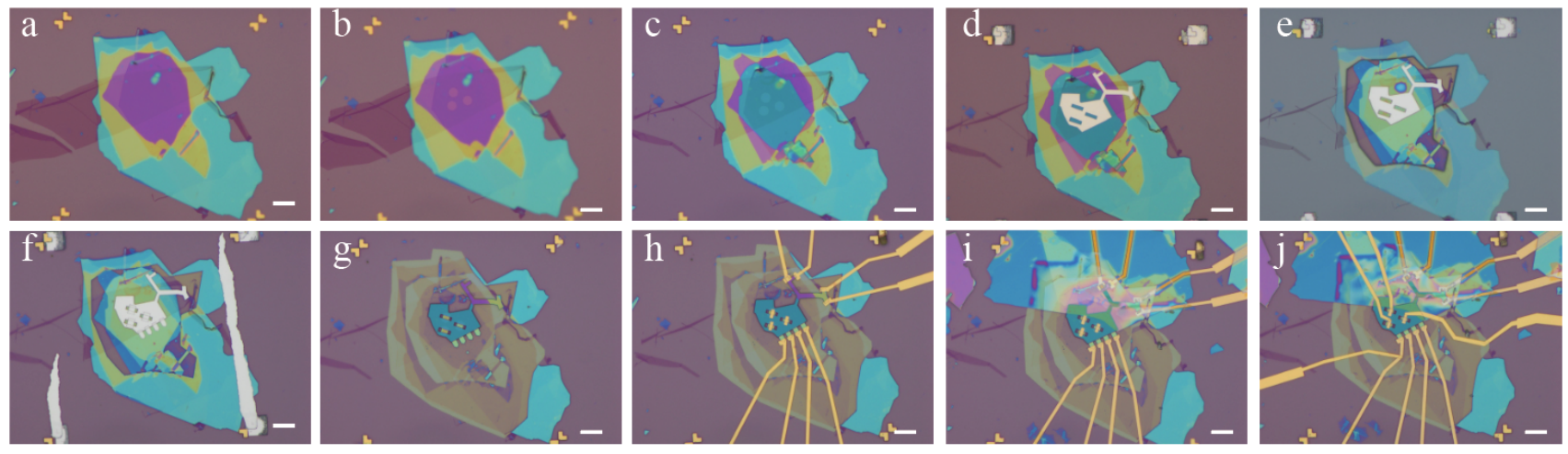}
\caption{\label{fig:SI_device} Optical images of the device at each fabrication step. \textbf{a.} The device is started with a stack consisting of graphite top and bottom gates, two layers of graphene, with BN in between them. 
\textbf{b.} Circular openings are made in the graphite top gate. 
\textbf{c.} An hBN flake is transferred onto the top gate. 
\textbf{d.} An aluminum mask for defining the shape of the stack and slots for Corbino contacts is evaporated on top of the stack. \textbf{e.} The stack after the first etch. \textbf{f.} After evaporation of the second aluminum mask for contacts to the bottom graphene. \textbf{g.} The stack after the second etch. \textbf{h.} First evaporation of the contacts. \textbf{i.} After putting another BN on top. \textbf{j.} Second evaporation of contacts, leaving a completed device. The scale bar is 10$\mu$m.}
\end{figure*}

The stack is made using polypropylene carbonate (PC) film to pick up graphite top gate, two graphene layers, and the bottom gate in sequence, with BN around 40nm in between each conducting layer. Fig.~\ref{fig:SI_device} illustrates the different steps of the fabrication process. The detailed description of each step is as follows:
\begin{enumerate}
    \item[a.] We start by making openings on the graphite top gate using O$_2$ plasma(RIE, 60W, 300mT). Then a layer of BN is transferred on top of the stack to cover the openings.
    \item[b.] We then evaporate an aluminum mask to define the shape of the top graphene layer and the Corbino contacts.
    \item[c.] We use CHF$_3$/O$_2$ plasma(40/4sccm, 0.5Pa, 200W source power and 30W bias power) and O$_2$ plasma alternatively to remove the top two BN layers and the top gate. The etch rate is carefully calibrated so the BN beneath the top graphene is etched by only 5-10nm, preventing electrical short of the two graphene layers. 
    \item[d.] A second aluminum mask is evaporated on top of the first mask to define the contacts of the bottom graphene, and a subsequent CHF$_3$/O$_2$ etch, which etches through the entire stack, is performed.
    \item[e.] The contacts (Cr/Pd/Au=2nm/15nm/150nm) are evaporated in two steps: first, we make contacts to the internal slots on the top graphene layer; then another BN is transfered to cover the edge of the stack so the internal contacts can be connected to the leads; finally a deposition is performed to make all the other contacts. 
\end{enumerate}

\section{Calibration of capacitance lever arm}
\begin{figure*}
\includegraphics{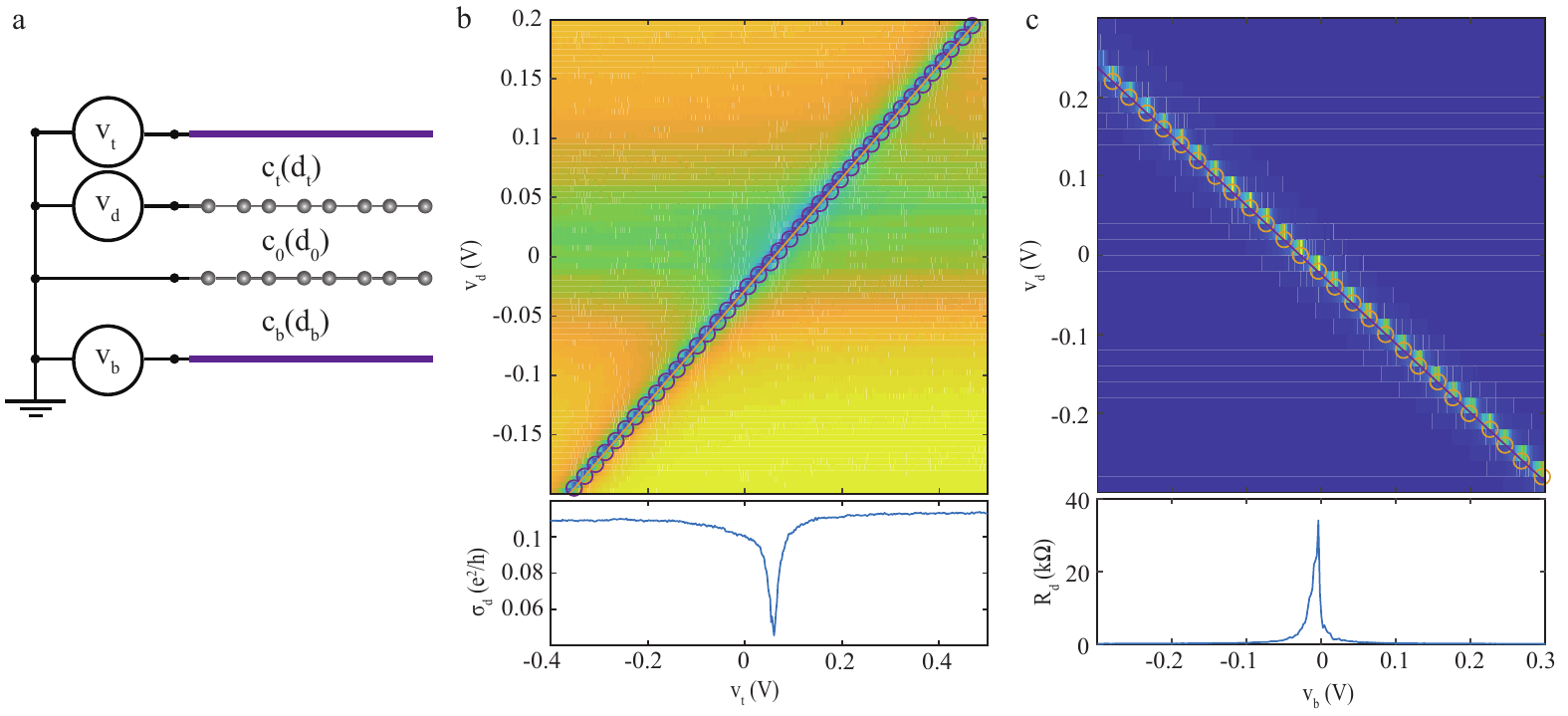}
\caption{\label{fig:SI_CapLeverArm} Determining capacitance lever arm by sweeping dual-gates. \textbf{a.} Measurement circuit. \textbf{b.} Top graphene conductance $\sigma_d$ as a function of $v_{t}$ and $v_{d}$. \textbf{c.} Bottom graphene resistance $R_d$ as a function of $v_{b}$ and $v_{d}$.}
\end{figure*}

The BN thicknesses determined from atomic force microscopy (AFM) measurement are $d_t$=45nm (for BN1 between top gate and top graphene), $d_0$=40nm (for BN2 between two graphene), and $d_b$=44nm (BN3 between bottom graphene and bottom gate); together, these in principle can be used to determine all capacitive lever arms.  However, the lever arm can be determined more accurately by measuring ratios of these capacitances directly \textit{in situ} by sweeping gate voltages and tracking the charge neutral point (CNP) of the graphene layers (Fig.~\ref{fig:SI_CapLeverArm}). To determine $d_0/d_t$, the bottom gate voltage is ramped according to $v_b= -v_dc_0/c_b$ to keep the bottom graphene density fixed. The carrier density in the top graphene is determined by $n_d=c_tv_t-(c_t+c_0)v_d$. At the CNP $n_d=0$, and therefore $\frac{v_d}{v_t}=\frac{c_t}{c_t+c_0}=\frac{1}{1+d_t/d_0}$. The slope of linear fit at CNP gives $v_d/v_t=0.475\pm.00023$(Fig.~\ref{fig:SI_CapLeverArm}b), which corresponds to $d_0/d_t=0.905\pm.00083$. Similarly, we can sweep $v_d$ and $v_b$ to determine $d_0/d_b$. The top gate voltage is set to $v_t=(1+c_0/c_t)v_d$ to keep the top graphene carrier density fixed. At CNP of the bottom graphene, $v_d/v_b=c_b/c_0=d_0/d_b=-0.8732\pm0.0015$(Fig.~\ref{fig:SI_CapLeverArm}c). 

\section{Sublattice splitting $\Delta_{AB}$}
\begin{figure*}[ht!]
\includegraphics{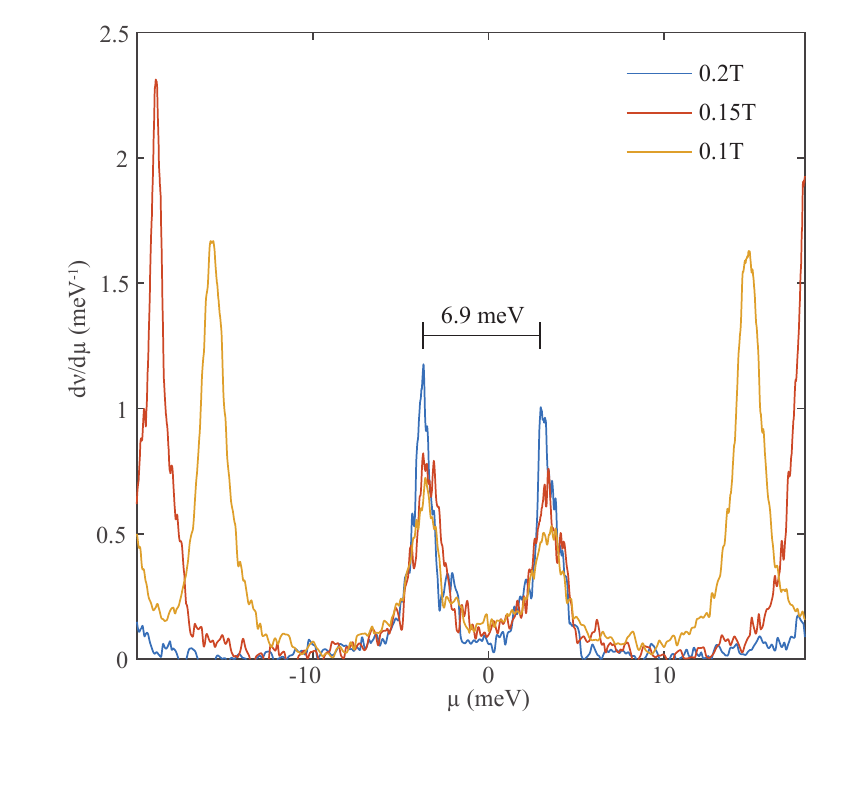}
\caption{\label{fig:SI_LFDOS} Density of states as a function of chemical potential at various magnetic fields.}
\end{figure*}

Fig.~\ref{fig:SI_LFDOS} shows $d\nu/d\mu\propto DOS$ at different magnetic fields. While the neighbouring cyclotron gaps are shifting with varying magnetic field, the gap at the charge neutral point is clearly independent of the magnetic field. Such a feature is consistent with a single particle AB sublattice splitting due to the Moir\'{e} superlattice between the graphene and the BN\cite{hunt_massive_2013,amet_composite_2015}.

\section{Fermi velocity normalization at zero magnetic field}
Here we give details about the determination of the Fermi velocity shown in Fig.~\ref{fig:figure1}f.
The correlation-induced renormalized Fermi velocity is described by the following equation\cite{gonzalez_marginal-fermi-liquid_1999,das_sarma_many-body_2007,polini_graphene_2007}:
\begin{equation}
    \frac{v_F}{v_F^0}=1-\frac{r_s}{\pi}[\frac{5}{3}+ln(r_s)]+\frac{r_s}{8}ln(\frac{n_c}{n})
\end{equation}
with $v_F^0=10^6$m/s being the single particle Fermi velocity and $n$ is the carrier density of the sample graphene. There are two fitting parameters: the interaction parameter $r_s=.437\pm.004$, and the ultraviolet cutoff $n_c=(.87\pm.02)\times 10^{14}$cm$^{-2}$.

\section{Transport in the detector graphene at finite magnetic field}
\begin{figure*}[ht!]
\includegraphics{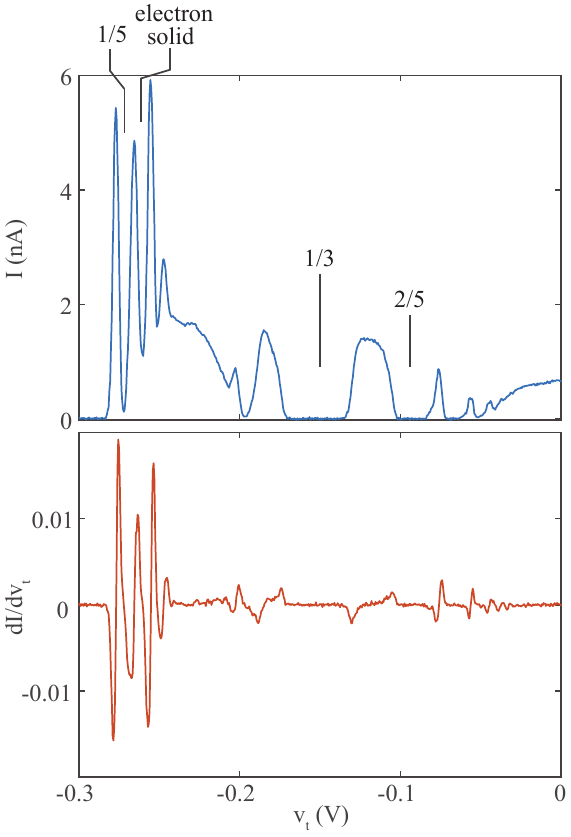}
\caption{\label{fig:SI_Corbino}
Transport measurement of the ``detector'' graphene at 14T and 15mK. The upper panel is the current through the graphene, and the lower panel is the top gate modulation of the current. }.
\end{figure*}

At high magnetic field, we keep the carrier density of the detector graphene fixed at a fractional quantum Hall gap. Most of our measurements are performed with the density fixed at $\nu=1/5$, where the local minimum is the sharpest (Fig.~\ref{fig:SI_Corbino}).

\section{Non-equilibrium state in the integer quantum Hall gap}
\begin{figure*}[ht!]
\includegraphics{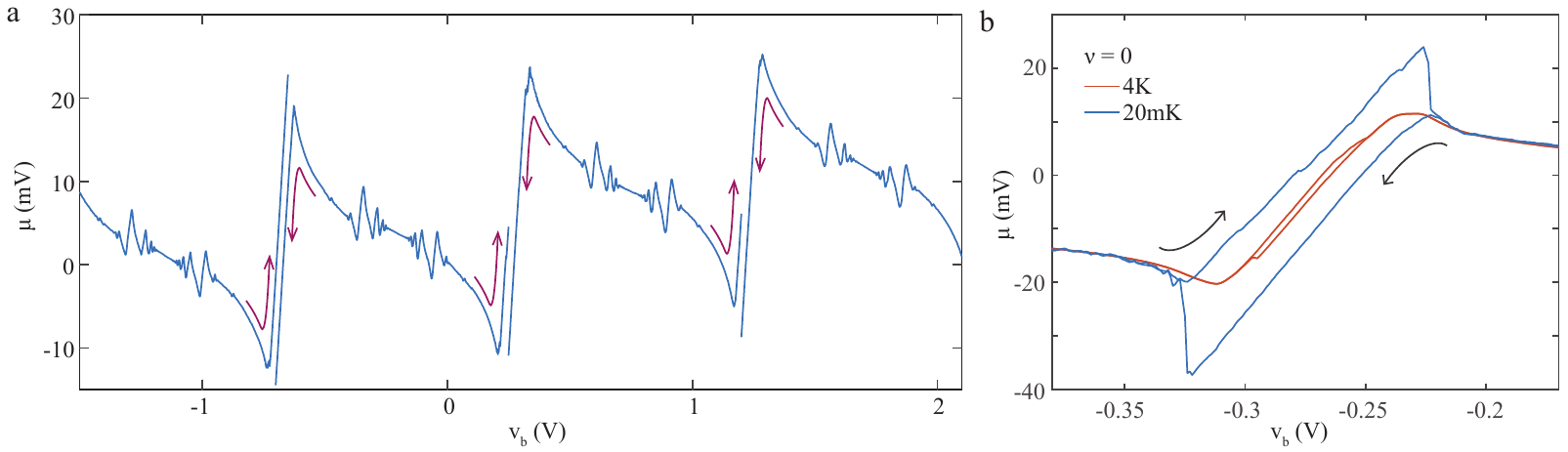}
\caption{\label{fig:SI_hysteresis}
Hysteresis in the integer quantum Hall gaps. \textbf{a.} $\mu$ as a function of $v_{b}$ in the N=0 LL. The data are taken at 14T and nominal 15mK. The red arrows mark the $v_{b}$ sweep direction. 
\textbf{b.} Detail of the $\nu=0$ integer quantum Hall gap. The arrows label $v_{b}$ sweep direction. The hysteresis is significantly reduced at 4K.}
\end{figure*}

As shown in Fig. \ref{fig:SI_hysteresis}, the chemical potential around the integer quantum Hall gaps within the ZLL shows hysteretic behavior when sweeping $v_{b}$ in opposite directions. The hysteresis is reduced at lower B, as well as at higher T, is nearly gone at 4K and B=14T. This phenomena has also been observed in GaAs 2DEG in several physical quantities, such as resistance\cite{zhu_hysteresis_2000,tutuc_layer-charge_2003,pan_hysteresis_2005,misra_dynamics_2008}, magnetization\cite{usher_magnetometry_2009,ruhe_origin_2009}, chemical potential\cite{ho_origin_2010}, and surface acoustic wave measurements\cite{pollanen_charge_2016}. 
% The phenomena has been attributed to the existence of non-equilibrium states inside the quantum Hall gap, but so far there has been no consensus as to the microscopic picture of the non-equilibrium states. One possible explanation attributes the non-equilibrium state to the edge states caused by eddy current; other model explains the non-equilibrium states to the charge transfer with some external charge reservoir (such as ionized impurities) when the carrier density inside the 2DEG bulk is extremely low. 
These nonequilibrium effects preclude measurement of the chemical potential in the regimes where they are observed.  All data presented in the main text are measured in the regime where no hysteresis is observed. 

\section{Comparison of the theoretically calculated $\nu=0$ gap with the experimental value}
\begin{figure*}[ht!]
\includegraphics{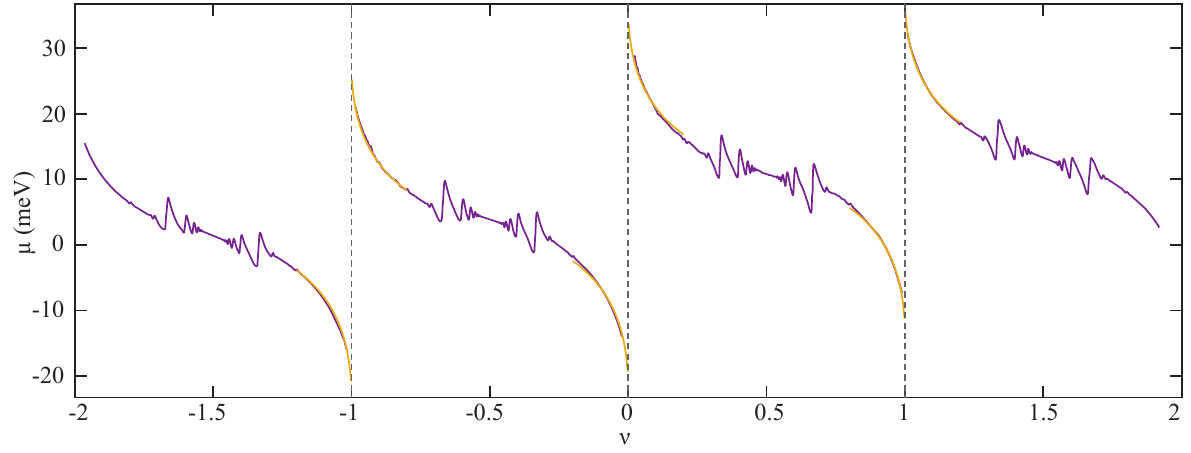}
\caption{\label{fig:SI_symmetryBreakingGaps}
$\mu(\nu)$ in the ZLL at 18T. The temperature is nominally 40mK. The data are extended to $\nu^*=0$ by using the Wigner crystal model described in the main text (orange curves).}.
\end{figure*}
Theoretically, the gap at $\nu = 0$ is predicted to be  $\Delta \mu = \int dq^2 V_{\textrm{scr}}(q) e^{-q^2 \ell_B^2/2} + \Delta_{AB} - E_Z = 55.3$meV, where we have used the screened Coulomb interaction $V_{\textrm{scr}}$ calibrated form Fig.~2b. $\int dq^2 V_{\textrm{scr}}(q) e^{-q^2 \ell_B^2/2} = 2(E(0)-E(1)) = 50.4$meV, where $E(0)$ and $E(1)$ are energy at $\nu^*=0$ and $\nu^*=1$ calculated by iDMRG (plot in Fig.~\ref{fig:figure2}c); $\Delta_{AB}=6.9$meV is the sublattice splitting; $E_Z\approx2$meV is the Zeeman energy. To compare with experiment, we extrapolate the $\mu(\nu)$ data,  which is cutoff in the window $|\nu| < 0.025$ due to the large IQHE charging time, to $\nu = 0$ using the Wigner-crystal model of Fig.~2b, giving a gap of $\Delta \mu = 53.1$meV, in very good agreement (4\%) with theory. Note that the \textit{bare} Coulomb interaction predicts  $\Delta \mu = 72.5$meV, supporting the importance of screening.

\section{Effective interaction from RPA and gate screening}
Here we present our model for calculating the dielectric function in Fig.~\ref{fig:figure2} in the main text, which takes into account screening from the proximate graphite gates and RPA screening from the Dirac sea of the graphene itself. The RPA treatment is adopted from Ref.~\cite{shizuya_electromagnetic_2007}.

The static dielectric function due to inter-LL virtual excitations of the MLG can be obtained within the random phase approximation (RPA):
\begin{align}
    \epsilon_\nu(q) &=1-V_0(q) {\Pi}_\nu(q,\omega=0) \\
    V_{\textrm{RPA}}(q) &= V_0(q) / \epsilon_\nu(q)
\end{align}
where $q$ is the wave vector  and ${\Pi}_\nu(q,\omega)$ is the polarizability of non-interacting graphene at filling factor $\nu$ and frequency $\omega$ (we ignore retardation effects by making the static approximation $\omega = 0$).
In the absence of gates, $V_0(q)$ would take the pure Coulomb form $V_0(q) = \frac{2{\pi}e^2}{\epsilon_{\mathrm{hBN}}q}$, with $\epsilon_{\mathrm{hBN}} = \sqrt{\epsilon_{\mathrm{hBN},\perp} \epsilon_{\mathrm{hBN},\parallel}}$ the dielectric constant of the surrounding boron-nitride substrate and $e$ the electron charge.
However, we also need to account for screening from the graphite gates, which we model as metallic equipotentials at distances $d_1, d_2$ below / above the graphene layer.
A standard electrostatic calculation shows that $V_0(q) = \frac{2{\pi}e^2}{{\epsilon_\mathrm{hBN}}q} f_{d_1,d_2}(q)$ for the form factor
\begin{equation}
    f_{d_1,d_2}(q)=2\frac{\tanh(\beta d_1 q ) \tanh( \beta d_2 q)}{ \tanh(\beta d_1 q)+\tanh(\beta d_2 q)}, \quad \beta = \sqrt{\frac{\epsilon_{\mathrm{hBN},\parallel}}{ \epsilon_{\mathrm{hBN},\perp}}}
\end{equation}
% This reduces to $f_{d}(q)= \tanh(\beta d q)$ in the approximation that $d_1=d_2=d$, where the gate screened interaction is
% \begin{equation}
%     V_0(q)=\frac{2{\pi}e^2}{{\epsilon_\mathrm{hBN}}q}\tanh( \beta d q)
% \end{equation}
% While the experiment is close to this regime, throughout we nevertheless use the values $d_1 = d_2 = 40nm$ inferred from AFM measures, though we find our results are not particularly sensitive to such details since $d_i / \ell_B \gg 1$.

\begin{figure}
\includegraphics{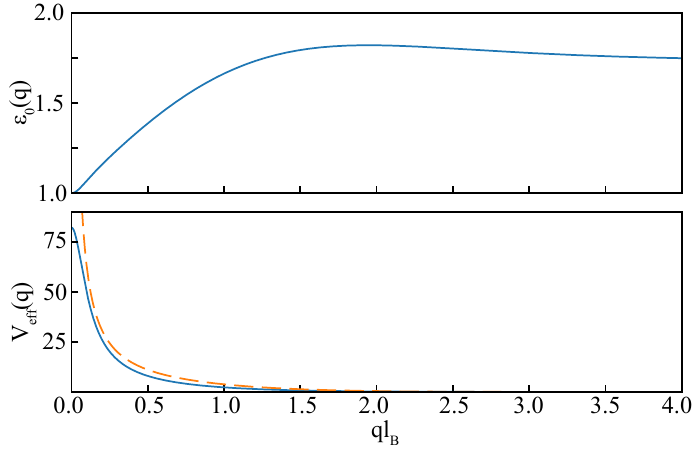}
\caption{\label{fig:SI_RPA}Dielectric function and effective interaction potential of Wigner crystal in the N=0 LL. Upper panel: Static dielectric function with gate screening ($\epsilon_{\mathrm{hBN}}=4.0$) and the RPA screening ($\alpha_G=1.85$) taken into account. Lower panel: Effective potential of the Wigner crystal. The orange dashed line is the effective energy without considering any screening effect, or $V_\mathrm{eff}(q)=\frac{2{\pi}e^2}{{\epsilon_{\mathrm{hBN}}}q}e^{-(ql_B)^2/2}$; the blue line is the effective potential with gate and RPA screening correction.}
\end{figure}

The polarizability (per isospin) consists of a sum over all inter-LL transitions $m \to n$ allowed by the Pauli principle:
\begin{align}
\Pi_\nu(q) = \sum_{ -\Lambda < m, n < \Lambda}  \nu_m (1 - \nu_n) \Pi_{m, n}(q)
\end{align}
where $m, n$ label LLs, $\nu_m$ is the filling of LL $m$, and $\Lambda \gg 1$ is a high energy cutoff. 
The contribution from each transition $\Pi_{m, n}(q)$ is sensitive to the structure of the LL wave functions via their ``form factors,'' as described in detail in Ref.~\cite{shizuya_electromagnetic_2007}.
For general $q$, the resulting sum must be evaluated numerically.
The result converges slowly with the cutoff (as $\Lambda^{-1/2}$), so we scale the cutoff from $\Lambda = 100 \to 200$ and extrapolate $\Lambda \to \infty$ with a quadratic polynomial in $\Lambda^{-1/2}$.
Calculating $\Pi_\nu(q)$ on a high-resolution grid ($\Delta q = 0.01 \ell_B^{-1}$), the result is then interpolated to continuous $q$ for input to the Wigner crystal and DMRG calculations.
The contribution to $\Pi$ from each of the four isospins is additive.

\section{Wigner crystal model}
The energy per electron of a classical Wigner crystal interacting through effective interaction $V_{\textrm{eff}}(r)$ is 
\begin{align}
\frac{N_e}{N_\Phi} E &= \frac{1}{2} \left(  \sum_{\mathbf{R}_i \neq 0} V_{\textrm{eff}}(R_i)  - \int d^2 r V_{\textrm{eff}}(r) \frac{\nu^\ast}{2 \pi \ell^2_B}  \right)  
= \frac{1}{2} \left( \frac{\nu^\ast}{2 \pi} \sum_{\mathbf{G}_i \neq 0} V_{\textrm{eff}}(\mathbf{G}_i) - V_{\textrm{eff}}(R_i = 0) \right)
\end{align}
Here $\mathbf{R}_i$ runs over the real-space Bravais lattice of the crystal, the exclusion $R_i \neq 0$ drops the self-interaction of the electron, and the subtraction accounts for the interaction between each electron and a neutralizing background charge density $\frac{\nu^\ast}{2 \pi \ell^2_B}$. Alternatively, it can be expressed as a sum over reciprocal vectors $\mathbf{G}_i$.
$E$ is the the energy per flux, so that $\mu(\nu) = \partial_\nu E(\nu)$ gives the desired chemical potential. 
Note that because of the background subtraction, $E < 0$, because the correlations of the Wigner crystal reduce the Coulomb interaction relative to a ``jellium'' of uniform charge. 

\begin{figure}[ht!]
\includegraphics{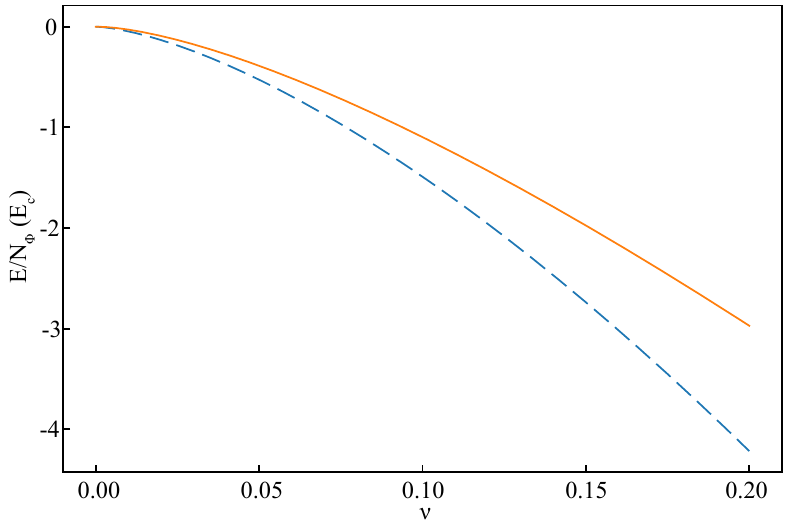}
\caption{\label{fig:SI_WCEnergy} Energy per flux. The orange dashed line plots the Wigner crystal energy calculated from bare Coulomb interaction\cite{levesque_crystallization_1984}; the blue line is the energy with gate and RPA screening taken into account ($\epsilon_{\mathrm{hBN}}=4.0$, $\alpha_G=1.85$).}
\end{figure}

For the effective interaction, we take the gate and RPA screened interaction discussed above and include in addition the ``form-factor'' $F_N(q)$ of the N-th LL: $V_{\textrm{eff}}(q) = V_{RPA}(q) |F_N(q)|^2$. In the N=0 LL, $F_0(q) = e^{-\frac{1}{4} (q \ell_B)^2} $.
The result can then be numerically evaluated in $q$-space, taking advantage of the form factor to cutoff the sum over $\mathbf{G}_i$ when $\mathbf{G}_i \ell_B \gg 1$.
The resulting energy, for both the bare and screened Coulomb interactions, is shown in  Fig.~\ref{fig:SI_WCEnergy}.

The treatment of the crystal as classical is valid so long as the wave functions of the electrons, which go as $|\phi(r - R_i)|^2 \propto e^{-\frac{r^2}{2 \ell_B^2}}$, are non-overlapping. This requires the interparticle distance satisfy $R \gg \ell_B$, or $\nu \ll 1$. 
At higher densities, the wave functions overlap and exchange-energy becomes important. However, at these higher densities the Wigner crystal melts and the electrons enter FQH states.

\section{Anisotropies in the N=1 Landau Level}
When taking into account only the long range ($r \gg a$) part of the Coulomb interaction, the N=1 LL has an SU(4) symmetry relating valley and spin.
However, lattice-scale effects (including the short-range part of the Coulomb interaction and phonons) break the valleys' involvement in this symmetry at order $a / \ell_B$, where $\ell_B$ is the magnetic length. 
The resulting ``valley anisotropies'' determine the nature of quantum-Hall symmetry breaking, as has been well explored both theoretically and experimentally in the N=0 LL\cite{alicea_graphene_2006,herbut_theory_2007,jung_theory_2009,nomura_field-induced_2009,kharitonov_phase_2012,khveshchenko_magnetic-field-induced_2001,young_spin_2012,young_tunable_2014}. 

In the N=0 LL, the  interaction anisotropy is thought to be well approximated by ``contact'' interactions ($V_{ani} \propto \delta(r)$). 
However, the N=0 LL is distinguished by the special form of its LL orbitals, which lock the valley and sublattice (A/B) degrees of freedom. In the N=1 LL, in contrast, both valleys are delocalized 50-50 over the two sublattices,  differing only in the precise shape (form factor) of their wave functions. 
Here we argue that this generically leads to anisotropies which are \textit{derivatives} of $\delta(r)$, leading to a very different density dependence in the FQH regime.

To understand the valley anisotropies in the MLG N=1 LL at a phenomenological level, it should be sufficient to consider the interaction arising from a short-range Hubbard-$U$ type interaction. The continuum field operator for spin $s$, sublattice $a$ is expanded in valleys $\tau$ as
\begin{align}
\hat{\psi}_{s a}(r) &= \sum_\tau e^{i \tau^z K \cdot r} \hat{\psi}_{s \tau a}(r) \\
&= \sum_{\tau, N, k } e^{i \tau^z K \cdot r} \phi^{\tau N k}_{a} (r)   \hat{c}_{s \tau N k} 
\end{align}
In the second line, we further expand the continuum operator in terms of Landau-gauge wave functions $\phi^{\tau N k}_{a} (r)$, where $k$ labels the Landau-gauge momenta and $N$ the LL index.
Henceforth, we restrict to the N=1 LL, so drop $N$ from the sum.
The N=1 LL-projected density operator for sublattice $a$ is then
\begin{align}
n_a(r) &= \sum e^{-i (\tau - \tau') K \cdot r} \bar{\phi}^{\tau k}_{a} (r) \phi^{\tau' k'}_{a} (r)   \hat{c}^\dagger_{s \tau k} \hat{c}_{s \tau' k'}    
\end{align}
It is then convenient to pass to momentum space using the technology of LL form-factors.
The MLG LL wave functions can be expanded as $ \phi^{\tau k}_{a} (r)  = \sum_n \phi^{\tau}_{n, a} \braket{r | n, k}$, where $\ket{n, k}$ is a Landau-gauge wave function of the $n$-th massive (GaAs-like) LL.
Inserting into the expression for $n_a(r)$ and Fourier transforming, the sublattice-resolved density operators are given in terms of GaAs form factors $F_{n, n'}$ and guiding-center operators $\bar{\rho}_{\mu \nu}(q) $.
Recall that the guiding-center operators between isospin components $\mu = (s, \tau), \nu$ are defined to be
\begin{align}
\bar{\rho}_{\mu, \nu}(q) = \sum_{k} e^{-i q_x k \ell_B^2} c^\dagger_{\mu\, k-q_y/2} c_{\nu\,  k + q_y/2}  
\end{align}
Inserting $n_a(r)$ into a Fourier transform, the density decouples into intra-valley and inter-valley contributions,
\begin{align}
n_a(q) &=  \sum_\tau \bar{\phi}^\tau_{n, a} \phi^\tau_{n', a}  F_{n, n'}(q) \bar{\rho}_{s \tau, s \tau} \\
n^{+}_a(q) &=  \bar{\phi}^{+}_{n, a} \phi^-_{n', a}  F_{n, n'}(q) \bar{\rho}_{s +, s -} \\
n^{-}_a(q) &=  \bar{\phi}^{-}_{n, a} \phi^+_{n', a}  F_{n, n'}(q) \bar{\rho}_{s -, s +} 
\end{align}
corresponding to the $q \sim 0$ and  $q \sim \pm 2 K$ parts of the density respectively. 
In the N=1 LL (ignoring the small mass $\Delta_{AB}$), $\phi^+ = (\ket{1}, \ket{0})/\sqrt{2}, \phi^- = (\ket{0}, -\ket{1}) /\sqrt{2}$.
The form factors are $F_{n, n} = e^{- q^2 \ell_B^2 / 4} L_n(q^2 \ell_B^2 / 2)$, where $L_n$ is the $n$-th Laguerre polynomial.

The most general form of a density-density interaction is then
\begin{align}
H &= \frac{1}{2} \sum_a \left[ V_{ab}(q) n_a(-q)  n_b(q)  + V_{ab}^+(q) n^{+}_a(-q) n^{-}_b(q) +  V_{ab}^-(q) n^{-}_a(-q) n^{+}_b(q) \right]
\end{align}
subject to constraints of symmetry and hermiticity.

\subsection{Intra-valley Hubbard-$U$}
We first consider the intra-valley part of a  sublattice-diagonal interaction
\begin{align}
H &= \frac{1}{2} \sum_{q, a}  U(q) n_a(-q)  n_a(q) 
\end{align}
For a Hubbard-U interaction, for example, the normalization is implicitly $U(r) = U_0 a^2 \delta(r)$, where $a $ is the graphene lattice scale and $U_0 \sim \frac{e^2}{4 \pi \epsilon a}$, so $U(q) = U_0 a^2$. In units ofquantum Hall scales $E_C$ and $\ell_B$, $u(q) = \ell_B^{-2} U(\ell_B q) / E_C =  U_0 \frac{a^2}{\ell_B^2} \frac{\epsilon \ell_B}{e^2} = U_0 \frac{\epsilon a}{e^2}  \frac{a}{\ell_B}$.  

The form-factor contraction takes the form
\begin{align}
 \sum_a \bar{\phi}^\tau_{n_1, a} \phi^\tau_{n_2, a} \bar{\phi}^{\tau'}_{n'_1, a} \phi^{\tau'}_{n'_2, a} F_{n_1, n_2}(-q)  F_{n'_1, n'_2}(q) &= \delta_{\tau \tau'} ( F_{0, 0}(-q) F_{0, 0}(q)  + F_{1, 1}(-q) F_{1, 1}(q) ) / 4 \\
 & + \sigma^x_{\tau \tau'} ( F_{0, 0}(-q) F_{1, 1}(q)  + F_{1, 1}(-q) F_{0, 0}(q) ) / 4
\end{align}
This leads to sum and difference 
\begin{align}
F(q) & \equiv  ( F_{0, 0}(q) + F_{1, 1}(q) )/2 \\
F^z(q) & \equiv ( F_{0, 0}(q) - F_{1, 1}(q) )/2 = e^{-\frac{1}{4}  \ell_B^2 q^2}  (1 - (1 - \ell_B^2 q^2/2) )/2 = e^{-\frac{1}{4} \ell_B^2 q^2}  \ell_B^2 q^2 / 4\\
H_U &= \frac{1}{2} \sum_q U(q) \frac{1}{2} \left(  |F(q)|^2   \bar{\rho}_{s \tau, s \tau}(q) \bar{\rho}_{s' \tau', s' \tau'}(-q)  + |F^z(q)|^2   \sigma^z_{\tau, \tau} \sigma^z_{\tau', \tau'}  \bar{\rho}_{s \tau, s \tau}(q) \bar{\rho}_{s' \tau', s' \tau'}(-q) \right)\\
H_U &= \frac{1}{2} \sum_q U(q) \frac{1}{2} \left(  |F(q)|^2   \bar{\rho}(q) \bar{\rho}(-q)  + |F^z(q)|^2  \bar{\rho}^z(q) \bar{\rho}^z(-q) \right)
\end{align}
Here $\bar{\rho}^\mu = \mbox{Tr}(\bar{
\rho} \tau^\mu )$.
Plugging in $F^z$, the anisotropy is
\begin{align}
H^z_U &= \frac{1}{2} \sum_q  e^{-\frac{1}{2}  \ell_B^2 q^2} \frac{ \ell_B^4 q^4 U(q)}{32}   \bar{\rho}^z(q) \bar{\rho}^z(-q) 
\end{align}
The key observation is that $U(q) \to  \ell_B^4 q^4 U(q)$. So even if $U(q)$ is taken to be a contact interaction, the effective interaction is not.

It is instructive to compare this with the analogous calculation in the N=0 LL, where $\phi^+ = (\ket{0}, 0), \phi^- = (0, \ket{0})$ (valley-sublattice locking). Following the same calculation, we then find $F^z(q) \propto F_{0, 0}(q) = e^{-\frac{1}{2} q^2}$, so the interactions is a simple contact interaction. 

\subsection{Inter-valley Hubbard-$U$}
The $q \sim 2 K$ part of the sublattice-resolved density operators take the form
\begin{align}
n^+_A(q) &= \frac{1}{2} F_{01}(q) \bar{\rho}^+(q) \\
n^+_B(q) &= -\frac{1}{2} F_{10}(q) \bar{\rho}^+(q) \\
n^+_A(q) n^-_A(-q) + n^+_B(q) n^-_B(-q) &=  \frac{1}{2} |F_{01}(q)|^2 \bar{\rho}^+(q) \bar{\rho}^-(-q)
\end{align}
So, by a similar argument as the intra-valley part, we obtain
\begin{align}
H^{xy}_U &= \frac{1}{2} \sum_q    U(q) |F_{01}(q)|^2    \bar{\rho}^+(q) \bar{\rho}^-(-q) 
\end{align}
where $\tau^{\pm} = \tau^x \pm i \tau^y$.
Plugging in the form-factors,  $|F_{01}(q)|^2 \propto q^2$. So, in contrast to the $q^4$ $z$-anisotropy, the $xy$-anisotropy scales with $q^2$. 

\subsection{Phenomenological Hamiltonian}
Together, this motivates a phenomenological anisotropy Hamiltonian of the form
\begin{align}
H_{ani} = \frac{E_C}{2} \sum_q  e^{- \frac{1}{2}  \ell_B^2 q^2 }  ( g_z  \ell_B^4 q^4 \bar{\rho}^z(q) \bar{\rho}^z(-q) +   g_{xy} \ell_B^2 q^2   \bar{\rho}^+(q) \bar{\rho}^-(-q) )
\end{align}
in units of $\ell_B$ and $E_C$. The dimensionless coefficients $g$ are expected to be of order $a / \ell_B$.
Passing back to real-space, the $q^2, q^4$ dependence maps on to the $\ell_B^{2m} \nabla^{2m} \delta(r)$ form given in the main text. 

To implement these anisotropies numerically, we note that a potential $V(q)$ can be expanded in terms of the ``Haldane pseudopotentials'' as $V(q) = 2 \sum_m L_m(q^2)$ (note there seems to be some disagreement in the literature on factors of $2 \pi$).
We can use this to determine the following pseudopotential decompositions $\{V_m\}$ for $q^{2m}$: $1 \rightarrow \{ \frac{1}{2} \}, - q^2 \to \{- \frac{1}{2},  \frac{1}{2} \},  q^4 \rightarrow \{ 1, -2, 1 \}$.
These Haldane pseudopotentials are then contracted with the appropriate index structure in the $\tau^{x/y/z}$ space and added to the Hamiltonian for two-component iDMRG calculations.

\end{document}